# Simulation and optimization of HEMTs

Hesameddin Ilatikhameneh · Reza Ashrafi · Sina Khorasani*

School of Electrical Engineering, Sharif University of Technology, Tehran, Iran
Email: khorasani@sina.sharif.edu

**Abstract** We have developed a simulation system for nanoscale high-electron mobility transistors, in which the self-consistent solution of Poisson and Schrödinger equations is obtained with the finite element method. We solve the exact set of nonlinear differential equations to obtain electron wave function, electric potential distribution, electron density, Fermi surface energy and current density distribution in the whole body of the device. For more precision, local dependence of carrier mobility on the electric field distribution is considered. We furthermore compare the simulation to a recent experimental measurement and observe perfect agreement. We also propose a graded channel design to improve the transconductance and thereby the threshold frequency of the device.

## 1 Introdcution

Nowadays, ultra-high speed circuits are mostly based on heterojunction devices including Heterojunction Bipolar Transistors (HBTs) and High Electron Mobility Transistors (HEMTs). HEMT-based high speed Integrated Circuits (ICs) and millimeter-wave microwave ICs [1] need to be scaled down to small dimensions for higher performance. In 1994 for the first time [2] a monolithic HEMT IC design was presented, which incorporated active regulated self-bias. The operation frequency of monolithic ICs has recently been extended well into the millimeter-wave range [3,4]. An advanced design of a highly integrated transmitter and receiver Monolithic Millimeter-wave ICs (MMICs) was reported in 2005 [5], based on a commercial 0.15μm, 88GHz (183 GHz MAX) GaAs pHEMT MMIC process and characterized on both chip and system levels.

While millimeter-wave applications call for active devices with higher cut-off frequencies, HEMTs offer the high frequency solution due to their high frequency operation and large current driving capabilities. In this regard, AlGaN–GaN HEMTs are emerging as the promising candidates for the radio frequency and microwave frequency power amplifiers used in advanced wireless communication systems [6]. Single-heterojunction (SH) HEMTs employing one doping layer have shown excellent current gain cut-off frequencies and extremely low noise figurers [7]. However, they often suffer from low current densities due to the relatively small number of carriers in the channel. The sheet carrier density can be improved to some extent by increasing the doping in the donor layer at the expense of lower breakdown voltages. A better approach to achieve high current driving capability is by distributing the doping into multiple donor regions and employing multiple heterojunctions. In this way, multiple Two-Dimensional Electron Gases (2DEGs) are formed and a high current density can be expected [8]. High power composite channel GaInAs/InP HEMT [9] and dual-delta-doped power HEMTs [10] are typical instances of multiple heterojunction devices.

For many different circuits, design and applications of HEMTs, accurate models for various characteristics of the device are needed. In [11] the authors reported a new empirical and simple model that can represent the current–voltage (I–V) characteristics of HEMT devices with high accuracy. An inverse modeling technique was introduced in [12] to determine the structural and physical parameters of HEMT from the desired data for maximum transconductance. Presented analytical model by [13] for I-V characteristics of strained and lattice matched HEMTs on InP substrate using a variational charge control model has resulted in an accurate description of the device. This also includes correct modeling of the subthreshold and saturation regions, which have a particular importance for digital applications as well as the microwave power operation of HEMTs. In this approach, instead of linear capacitance approximation [14,15] a polynomial channel charge density versus gate-to-channel voltage relation has been adopted.

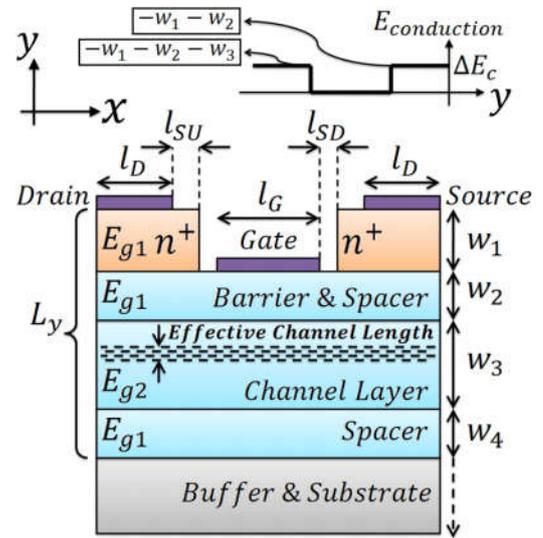

Fig. 1. Typical HEMT structure.

In general, for more reliable modeling and simulation of high speed and high performance nanoscale heterojunction devices, a self-consistent and accurate coupling of quantum mechanical and electrostatic analyses should be addressed. Additionally, various optimization approaches for improving the performance of HEMTs have been reported up to now. In this regard, a novel and accurate method for simulating nonequilibrium gate current and sheet carrier concentration in AlGaAs/GaAs HEMT structures has been reported in [16]. A first-principles theoretical comparison of the performance of identical $Al_{0.32}Ga_{0.68}As/GaAs$ and $Al_{0.15}Ga_{0.85}As/GaAs$ pseudomorphic HEMTs based on an ensemble Monte Carlo simulation coupled with a 2D Poisson solver has been published in [17]. A paper published in 1999 [18] presented a new and high-performance InGaP/In$_x$Ga$_{1-x}$As HEMT with an inverted delta-doped V-shaped channel. Due to the presence of V-shaped inverted delta-doped InGaP/In$_x$Ga$_{1-x}$As structure, good carrier confinement and a flat and wide transconductance operation regime could be expected.

Among notable works in the transport simulation of semiconductor devices [19], there has been a wide range of recent literature discussing quantum corrected Monte Carlo [20] and quantum corrected drift-diffusion models [21-24]. These days, advanced and expensive commercial softwares such as Silvaco [25] are able to fully incorporate quantum effects into the operation of semiconductor devices.

In this work, we solve the exact nonlinear coupled Schrödinger and electrostatic equations (Poisson and charge conservation), in 2D geometry, using the Finite Element Method (FEM). These equations are solved self-consistently in an iterative manner until the solution converges. As an example, we consider a GaN-based HEMT for determine a typical cut-off frequency of device. As a new approach for improvement of HEMT characteristics, we also have considered and simulated the device with various band edge energy profiles in channel layer, resulting from graded control over the Aluminum fraction. We show that it is possible to achieve a 53% improvement in the transconductance using an optimized profile of impurity in the channel layer. This study is based on the combination of Finite-Element solver FlexPDE 5 [26], which is designed for solution of partial differential equations, and MATLAB codes.

## 2 Theory

### A. Basic Equations

The quantum mechanical Schrödinger equation governing the distribution of electric charges is nonlinearly coupled with the Poisson and charge conservation equations, which are in turn dependent on the probabilistic wave functions. Hence, this system of equations must be solved self-consistently for a correct simulation of HEMT devices. Solution of the Schrödinger equation results in energy eigenvalues and electron wave functions, and thereby the electron density. Electric field and potential distribution are then extracted from the solution Poisson equation. Fermi surface distribution and current density of device are then obtained when charge conservation law is applied.

Here, we simulate the HEMT with two different approaches. In the first approach, the true distribution of electron current is neglected and Fermi energy surface is supposed to be constant, while in the second approach, the current continuity equation is applied along with the self-consistent system of Poisson-Schrödinger equations. Hence, the former is applicable to unbiased structures at equilibrium, the latter method may be used for analysis of fully biased configurations at unequilibrium. Although both methods are equally applicable to all typical HEMT structures based on III-V compounds, we choose the (AlGa)N family, which has drawn particular attraction in the recent years. The typical HEMT structure under consideration is shown in Fig. 1.

### B. Zero Bias

The first approach can be used for calculation of electron density distribution in the channel, and also derivation of the related parameters such as gate-bulk capacitance, when no drain-source voltage is applied. At the beginning of analysis we assume that all the carriers are attached to their associated dopant atoms, which implies zero charge density $\rho(x,y) = 0$ everywhere in the device. Hence, the initial electric potential can be estimated from the Laplace's equation

$$\nabla^2 V = -\frac{1}{\varepsilon}\rho \,, \qquad (1)$$

where $V$ is the 2D electric potential. This potential distribution is obtained regardless of the difference between the band-gap energies of semiconductors. Therefore, using the superposition law, the total potential energy $U_t$ can be obtained in the form of

$$U_t = qV - E_c \,, \qquad (2)$$

in which $q$ is the electronic charge and $E_c$ is the conduction band energy. The initial function profile of $E_c$ could be approximated with combination of smooth step functions formed with tangent hyperbolic functions as is shown in Fig. 2. Now, the Schrödinger equation can be written as

$$-\frac{\hbar^2}{2m}\nabla^2\psi(x,y) + U_t(x,y)\,\psi(x,y) = E\,\psi(x,y) \,, \quad (3)$$

where $\psi(x,y)$ is the electron wave function, $\hbar$ is the reduced Planck's constant, $m$ is the electron effective mass, and $E$ is the electron energy. Here, the total electrostatic potential energy $U_t(x,y)$ is simply substituted by $-qV_t$. By solving (3), the eigenvalues $E_i$ and eigen-functions $\psi_i$ of the $i$-th state could be obtained.

Now considering the fact that the energy states are discrete and do not form a continuum [27], the total electron density should calculated by applying the Fermi statistics to find the summation of electron densities in the energy sub-bands [28]. This results in

$$n(x,y) = \sum_{i=0}^{\infty}\left|\psi_i(x,y)\right|^2 n_i \,. \qquad (5)$$

Here $\psi_i(x,y)$ is the corresponding wave function of the $i$th energy state, and $n_i$ is the electron density at this specific energy state.

In the case of 2D analysis, the electron density for the $i$th energy state is obtained from the equation [29,30] (Appendix)

$$n_i = \frac{2mkT}{\pi\hbar^2}\ln\left[1 + \exp\left(\frac{E_f - E_i}{kT}\right)\right]. \qquad (6)$$

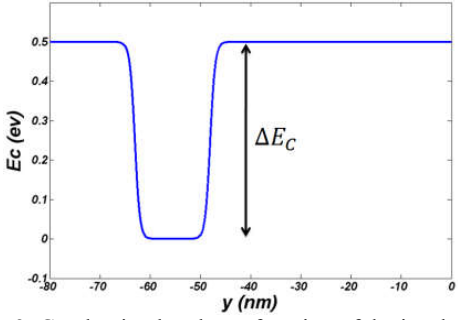

Fig. 2. Conduction band as a function of device depth.

where $k$ is the Boltzmann's constant, $T$ is the absolute temperature, and $E_f$ is the Fermi energy. The Fermi Energy $E_f$ is here obtained from the bulk doping density. Also, the conduction band of the channel layer $E_{c2}$ can be considered as the reference energy for the whole of the system, and the differences between the energy levels of $E_f$ and $E_c$ of the channel layer semiconductor is attained using

$$n = N_d = N_C \exp\left(\frac{E_f - E_i}{kT}\right). \qquad (7)$$

The typical resulting energy band diagram is illustrated in Fig. 3. Now, the 2D electron density $n(x,y)$ obtained from (5) can be iteratively plugged in the Poisson equation (1), resulting in

$$\nabla^2 V = \frac{-q\left[N_d - n(x,y)\right]}{\varepsilon} \quad . \qquad (8)$$

After solving (8), the total potential distribution is obtained using the new potential, and $-qV_t$ replaces $U_t(x,y)$ in the Schrödinger equation (3). Therefore, the energy states and eigen-functions are approximated using the new potential energy distribution. In this way, the iteration loop is closed and repeated until the electron density $n(x,y)$ converges to a steady distribution. A flowchart outlining the mentioned procedure is shown Fig. 4.

It should be noted that the Schrodinger equation is solved in 2D subject to the boundary conditions given in Table 1. This will typically result in a discrete spectrum, with each state being highly occupied due to in-plane momentum of carriers. This is because the confinement occurs mainly in *y*-direction, and these are treated by integration of energy states over normal directions by approximating the dense spectrum as a parabolic subband, which automatically result in (6). This has been elaborated in Appendix.

Furthermore, other nonideal phenomena such as polarization and strain-induced effects, which are quite typical in III-V heterostructures are here ignored. Such effects only find importance in optoelectronic devices where accurate transition energies must be known and light absorption or emission spectrum is sought. For instance, in the optimal design of GaN Light-Emitting Diodes [31] and strongly-coupled AlGaAs quantum dots [32], both polarization and strain effects must be calculated using 4×4 matrix techniques, perturbation method and envelope approximation. This will enable one to find accurate solutions for electrons, heavy- and light-holes as well as holes in the split-off band.

For other applications where transitions do not significantly take place and only one type of carrier is involved, such as the conducting interface modulator [33] and this work, these may be therefore safely ignored. The reason is that, here, we look for charge density of the 2D electron gas (2DEG), which is by (6) only logarithmically dependent on the values of eigen energies. This will also help us to justify the accuracy of numerical simulations, which is here observed and described in the next section.

### C.  Biased Configuration

The second approach for derivation of the basic parameters in HEMTs takes care of the potential difference between the drain and source electrodes of the device, and hence can be used for analysis of biased configurations. As a result, the current density distribution in the bulk of the transistor is attained. The only difference between the new situation at unequilibrium (in the presence of $V_{DS}$) and the previous form at equilibrium (in absence of $V_{DS}$) is that the Fermi energy level is no longer flat and constant throughout. Indeed, the Fermi energy needs to be replaced by the quasi-Fermi level.

The quasi-Fermi energy level at the source and drain contacts are equal to their respective applied voltages, while in other regions it will be derived from

$$\nabla \cdot \mathbf{J} = 0 \quad , \qquad (9)$$

where

$$\mathbf{J} = n\mu \nabla E_f \quad , \qquad (10)$$

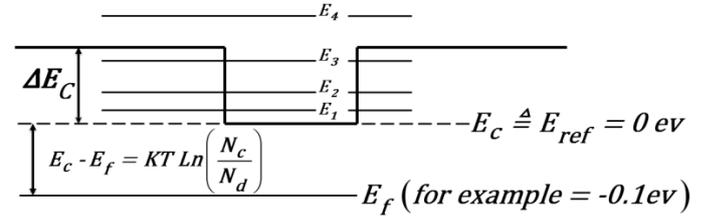

Fig. 3. Energy states in quantum well of channel layer.

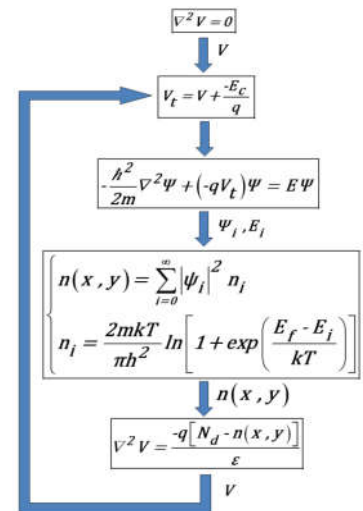

Fig. 4. Flowchart of first simulation approach algorithm.

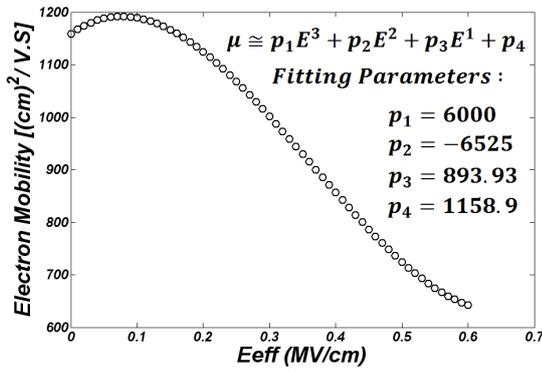

Fig. 5. Electric field dependency of GaN electron mobility, based on data taken from [23].

One can also consider the electric field dependence of electron mobility $\mu = \mu\left(E_{eff}\right)$ in GaN, where $E_{eff}$ is the effective electric field, through curve fitting to the available experimental data [34]. This is done according to the polynomial fit as shown in Fig. 5. Therefore, the mobility $\mu$ becomes a local function of coordinates, and the divergence condition (9) changes into

$$\nabla \cdot \left[\mu(x,y)\, n(x,y)\, \nabla E_f(x,y)\right] = 0 \quad , \tag{11}$$

In the above equation, $n\left(x,y\right)$ is replaced from (5), which is dependent on the Fermi energy. Finally, (11) recasts into the non-linear differential equation reading

$$\nabla \cdot \left\{\left[\sum_{i=0}^{\infty} \left|\psi_i\left(x,y\right)\right|^2 \ln\left(1 + \frac{E_f - E_i}{kT}\right)\right]\nabla E_f\right\} = 0 \quad . \tag{12}$$

By solving the above differential equation, the Fermi level energy distribution in the whole body of the transistor is determined. Therefore, $n\left(x,y\right)$ and the resulting current density are attained at each node separately. The flowchart used by the program is shown in Fig. 6. Since the second approach is more comprehensive and gives the complete I-V characteristics of the device, it needs significantly more computation.

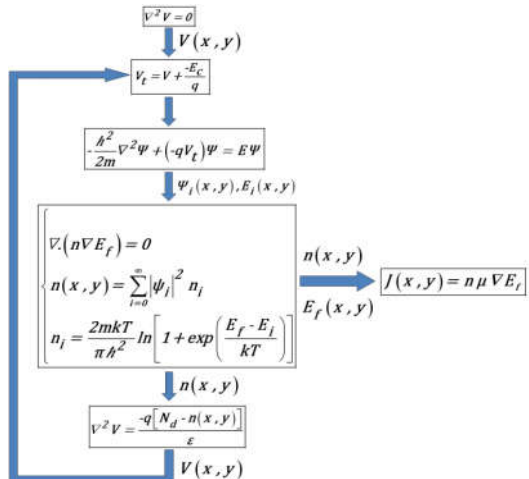

Fig. 6. Flowchart of second simulation approach algorithm.

## D. Boundary Conditions

The applied boundary conditions for the Schrödinger, Poisson and current continuity equations are shown in Fig. 7 and also reiterated in Table 1 for more clarity. The reader may notice that here the Schrödinger equation is solved in 2D at once, instead of only solving for the wavefunctions at the band edges in 1D across the quantum well.

Table 1. Boundary conditions in the simulation.

| Number | Details |
|---|---|
| 1 (Drain) | $V = V_d, \psi = 0, E_f = -qV_d + \varphi_{ms}$ |
| 2 (Gate) | $V = V_g, \psi = 0, E_f = -qV_g + \varphi_{ms}, \frac{d}{dn}E_f = 0$ |
| 3 (Source) | $V = V_s, \psi = 0, E_f = -qV_s + \varphi_{ms}$ |
| 4 | $\frac{d}{dn}V = 0, \psi = 0, \frac{d}{dn}E_f = 0$ |

Table 2. Dimensions of the simulated HEMT.

| $W_1$ | $W_2$ | $W_3$ | $W_4$ | $l_D$ | $l_G$ | $l_{SU}$ | $l_{SD}$ |
|---|---|---|---|---|---|---|---|
| 20nm | 28nm | 15nm | 10nm | 30nm | 20nm | 5nm | 5nm |

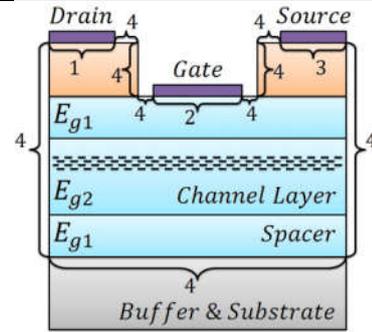

Fig. 7. Boundary conditions determination.

## 3 Results
### A. Electron Density

The dimensions of the double heterojunction device shown in Fig. 1 are enlisted in Table 2. All equations are solved using the standard FEM, and the solution converges quickly only after four cycles of repeating the outer loop in the flowchart, as shown in Fig. 6. Only the first four electron wave functions as plotted in Fig. 8 are taken into account. This is because of the fact that the first few eigen-functions are dominant in determination of overall electron density distribution, and the rest have negligible occupation and therefore are only of minor importance. After evaluating the electron wave function, we can calculate the electron density via (5) which is shown in Fig. 9. In addition, the electron density of a single heterojunction simulated HEMT with same dimension is plotted in the Fig. 10 for comparison purposes.

The electron density in the single heterojunction structure is more outspread in comparison with the double heterojunction structure. Obviously the corresponding peak electron density is lower, too. As it can be seen from Figs. 9 and 10, in the double heterojunction structure the maximum carrier density is approximately one order of magnitude larger than that of the single heterojunction structure; additionally, in the double heterojunction HEMT, electrons are more confined. This arises from the fact that electrons spread more uniformly over the channel layer.

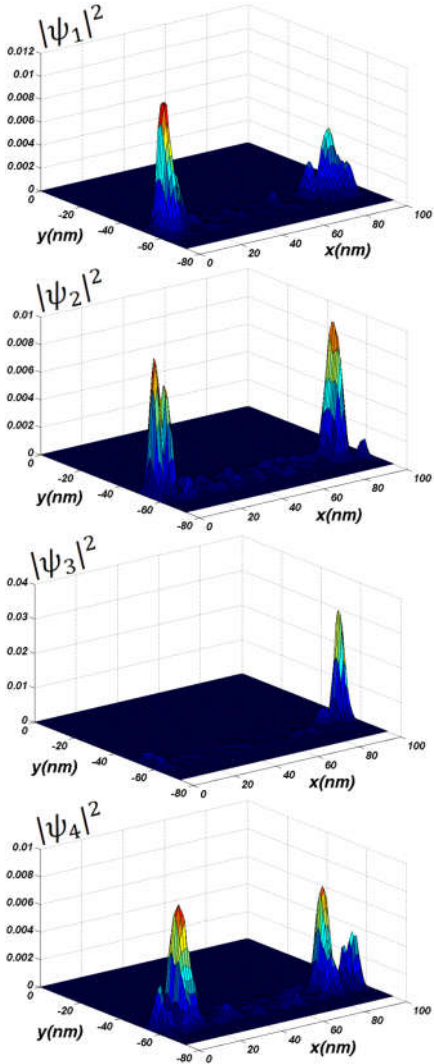

Fig. 8. The first four normalized electron wave functions of the double heterojunction device.

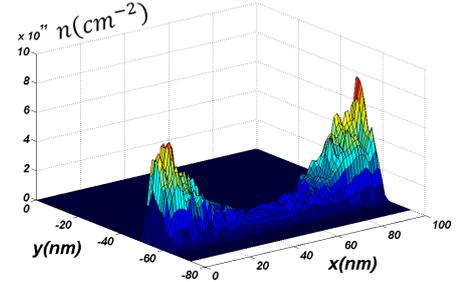

Fig. 9. Total electron density distribution in the double heterojunction device.

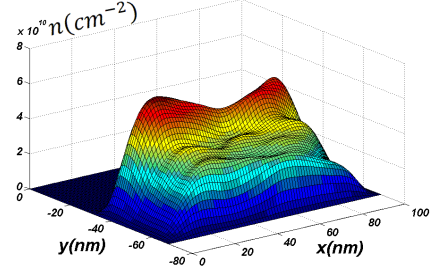

Fig. 10. Total electron density distribution in the single heterojunction device. In contrast to Fig. 9 for the double heterojunction structure, the density is much more uniform.

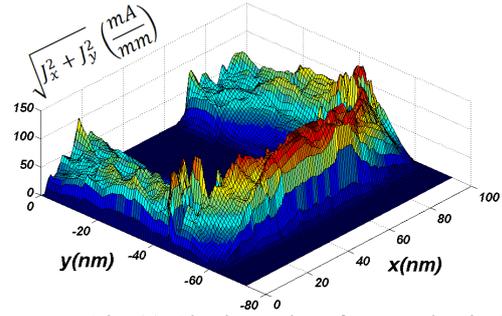

Fig. 11. Absolute value of current density in device.

The ground state is calculated for every bias point in 2D, which is naturally subject to asymmetric boundary conditions across the channel. Furthermore, the current does not flow completely horizontal as Fig. 11 clearly shows. While simple non-degenerate 1D quantum systems are known to have no nodes in their ground states (and the order of states is characterized by the number of zeros of wavefunctions), there is no reason that the same concept still could be expected under 2D and strongly asymmetric conditions.

As it is normally expected, the electron current path is observed to be through the channel layer. This effect is seen clearly in Fig. 11. Here, neglection of higher-order eigenfunctions results in a profile with locally rapid variations.

### B.  I-V Characteristics

The simulated DC I-V characteristics is shown in Fig. 12 and compared with the experimental result by Kwon *et al* [8]. A remarkable agreement between the numerical simulation and experimental results is observed. We can also estimate the cut-off frequency using the expression [35]

$$f_c = \frac{g_m}{2\pi C_{gs}} \ ,  \qquad (13)$$

where $g_m$ is the transconductance of the device. Also, the Gate-Source capacitance $C_{gs}$ can be obtained from Taylor series expansion [36] as

$$C_{gs} = C_{gs1} + C_{gs2}V_{gs} + C_{gs3}V_{gs}^2 . \qquad (14)$$

For a barrier thickness of 30nm, the coefficients of the above series for GaN HEMTs are [36] known to be $C_{gs1} = 0.473 \ pF/mm$ , $C_{gs2} = 0.018 \ pF/mmV$ , and $C_{gs3} = 0.006 \ pF/mmV^2$ . The transconductance is then calculated at $V_{gs} = 0.25V$ and $V_{ds} = 1V$ using the obtained $I_d - V_{ds}$ characteristics (Fig. 13). Calculated cut-off frequencies for double and single heterostructure are 189.5GHz and 171.1GHz, respectively.

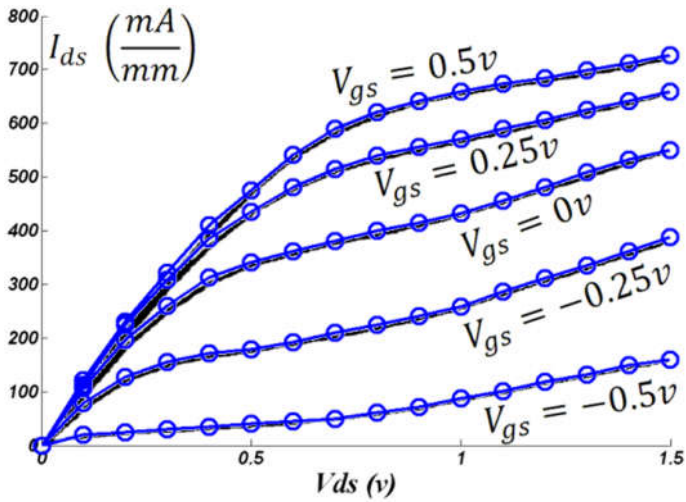

Fig. 12. I-V characteristic of the double heterostructure GaN HEMT: numerical simulations (blue lines) versus experimental data (black solid lines) from [8]. The agreement is very remarkable between the numerical simulation and experimental data and the fit is perfect.

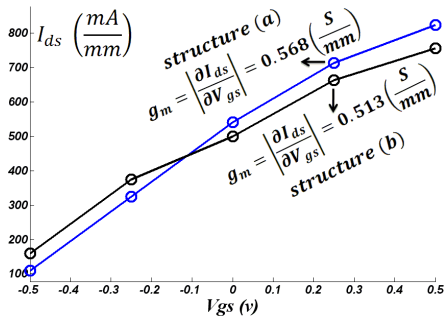

Fig. 13. Simulated the transconductance of HEMT at bias voltages $V_{gs} = 0.25V$ and $V_{ds} = 1V$: (a) Double Heterostructure; (b) Single Heterostructure.

## 4 A Novel Structure

In this section, we show that it would be possible to increase the transconductance of the HEMT transistor by more than 53%, through engineering the profile of Aluminum in the channel layer. We first notice that based on the results for single and double hetero junction HEMTs with the same dimensions, it was observed that double heterojunction HEMT has a larger transconductance. As a result, it enjoyed higher cut-off frequency, too. One of the causes for such an improvement is better carrier confinement in the channel, which is typical for double heterojunction devices. This feature is clearly demonstrated in Figs. 9 and 10, in which carriers have tighter confinement and higher concentration in channel layer for the double heterojunction structure. Hence, we can conclude that a proper design of potential well, and specifically, the profile of band edge energy, would contribute to a superior carrier density distribution and confinement.

Here, we have simulated and compared two structures with the same dimensions, but different conduction band edge energy profiles across the channel layer. The first simulated structure is the same double heterojunction HEMT illustrated in Fig. 1. Conduction band edge energy profiles of this standard structure due to a controlled Aluminum fraction and our proposed structure, along $y$-direction are respectively shown in Figs. 14 and 15.

We employ a simple linear interpolation for the bandgap dependency of the ternary compound $Al_xGa_{1-x}As$ [37], and estimate the Al fraction needed at each depth for constructing this structure has also been calculated and shown in Figs. 14, 15. It should be possible to realize such a graded profile of doping using layer-by-layer growth techniques, such as Molecular Beam Epitaxy (MBE) or Metal-Organic Chemical Vapor Deposition (MOCVD).

Referring to Fig. 1, the dimensions of these two simulated structures are given in Table 3. Through numerical simulation we find that the carrier confinement in the proposed structure is significantly higher than the conventional double heterojunction HEMT (see Fig. 16 and Fig. 17). Hence, the effective channel thickness is reduced roughly by a factor of 2.1; the effective thickness of channel is here defined as the length scale of decay of carrier density across the $y$-direction (see Fig. 18). Therefore, a channel with uniform carrier density and thickness $L_{eff}$ would support the same total number of carriers.

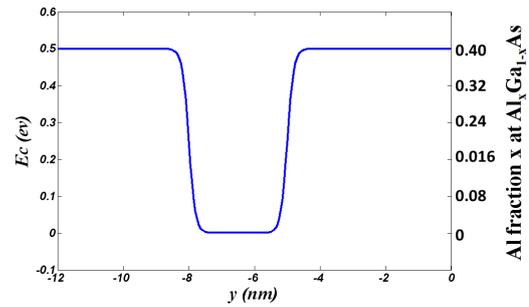

Fig. 14. Profile of conduction band energy and Al fraction of the simulated structure (the same double heterojunction HEMT that was investigated above just with smaller dimensions).

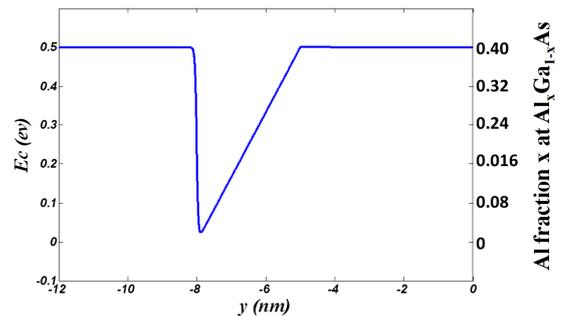

Fig. 15. Profile of conduction band energy and Al fraction of our suggested novel structure.

Table 3. Dimensions of the simulated and compared two HEMTs.

| $W_1$ | $W_2$ | $W_3$ | $W_4$ | $l_D$ | $l_G$ | $l_{SU}$ | $l_{SD}$ |
|---|---|---|---|---|---|---|---|
| 3nm | 2nm | 3nm | 4nm | 2nm | 2nm | 0.5nm | 0.5nm |

The effective channel thicknesses for the two compared structures are shown in Figs. 16 and 17, for bias voltages of $V_{gs} = 0V$ and $V_{ds} = 0.5V$. Calculated $I_{ds}$-$V_{gs}$ curve for $0 < V_{gs} < 0.35V$ and $V_{ds} = 0.5V$ are also calculated and plotted in Fig. 19. As it can be clearly seen here, our suggested structure has a higher transconductance due to the much better carrier confinement. Based on Fig. 19, the maximum enhancement in transconductance can be easily estimated to be about 53%.

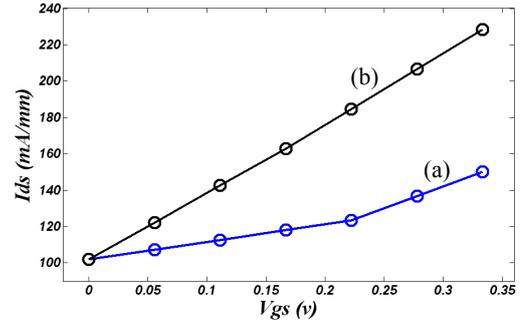

Fig. 19. Calculated transconductance of two compared HEMTs at bias voltages $V_{ds} = 0.5V$ and $0 < V_{gs} < 0.35V$ : (a) Double Heterostructure; (b) Our novel Heterostructure.

## 5  Conclusion

The FEM has been applied to simulation of GaN HEMTs. Two different approaches for simulation of HEMT devices, under equilibrium with zero bias, and at unequilibrium when biased, were reported. Poisson-Schrödinger equations were solved self-consistently until the solution converged. Through addition of the semi-classical current continuity equation, the effect of drain-source voltage could be considered. Obtained $I_{ds}$-$V_{ds}$ characteristics of HEMT devices from this simulation, was shown to be in complete agreement with the reported experimental results. Single and double heterojunction HEMT structures were compared and superior performance of double heterojunctions was confirmed by numerical simulations. Based on the simulation results, double heterojunction HEMTs had higher transconductance and therefore higher cut-off frequencies, which was a result of better confinement of electrons in the channel layer. Through engineered design of doping profile in the channel layer, the possibility of a significant enhancement in transconductance has been established.

## Appendix

The density of $i$-th energy state as given in (6) can be obtained by first noting that the 2D density of states of confined electrons is independent of their energy, given in energy or momentum spaces as

$$g_{2D}(E)\,dE = \frac{2m}{\pi\hbar^2}\,dE, \tag{A.1}$$

$$g_{2D}(k_\perp)\,d^2k_\perp = \frac{2}{(2\pi)^2}\,dE, \tag{A.2}$$

where the factor 2 is inserted to take account for the spin-degeneracy. This corresponds to assuming an isotropic confinement normal to the $y$-direction in $xz$-plane, with an electron energy for a parabolic band as

$$E(k_\perp) = E_i + \frac{1}{2m}\hbar^2 k_\perp{}^2. \tag{A.3}$$

Now, the electron density in the i-th subband can be found via integration as

$$n_i = \iint \frac{2}{(2\pi)^2} f\left[E\left(k_\perp\right); E_f\right] d^2 k_\perp$$

$$= \int_{E_i}^{\infty} g_{2D}\left(E\right) \frac{dE}{1 + \exp\left(\dfrac{E - E_f}{kT}\right)} \qquad (A4)$$

$$= \frac{2m}{\pi \hbar^2} kT \ln\left[1 + \exp\left(\dfrac{E_i - E_f}{kT}\right)\right],$$

in which $f\left(E; E_f\right)$ is the Fermi-Dirac distribution.

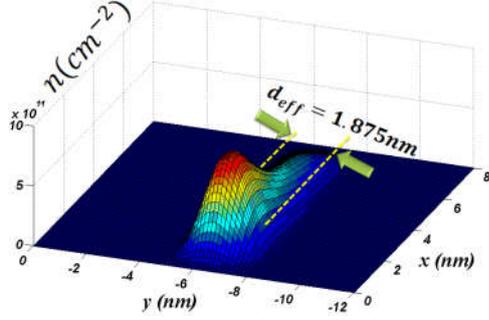

Fig. 16. Total electron density distribution in the double heterojunction with conduction band edge energy profile shown in Fig. 14. The effective channel thickness is calculated to be around 1.875nm.

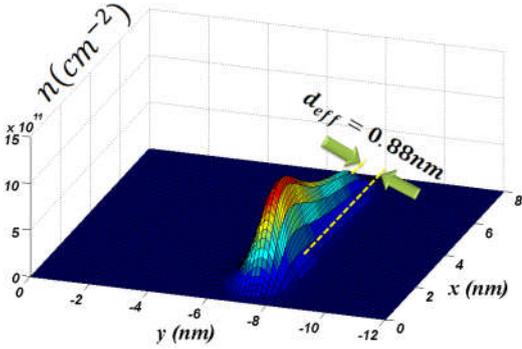

Fig. 17. Total electron density distribution in our suggested novel structure with conduction band edge energy profile shown in Fig.15. The effective channel thickness in this case is only 0.88nm.

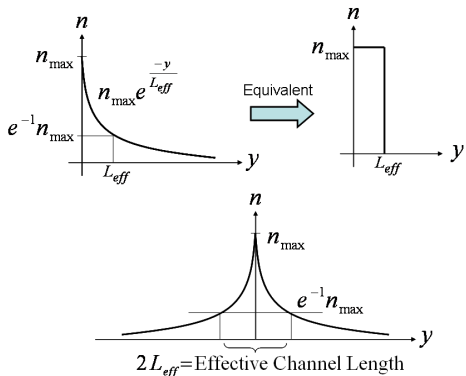

Fig. 18. Definition of criterion for calculation of effective channel length.